\documentclass[12pt]{article}
\usepackage{dina4p}
\usepackage{epsfig}

\def\Journal#1#2#3#4{{#1} {\bf #2}, #3 (#4)}

\def\NPB{{\em Nucl. Phys.} B}
\def\PLB{{\em Phys. Lett.}  B}
\def\PRL{\em Phys. Rev. Lett.}

\def\ZPC{{\em Z. Phys.} C}
\def\CPC{\em Computer Phys. Commun.}
\def\lsim{\mathrel{\rlap{\lower4pt\hbox{\hskip1pt$\sim$}}
    \raise1pt\hbox{$<$}}}                
\def\gsim{\mathrel{\rlap{\lower4pt\hbox{\hskip1pt$\sim$}}
    \raise1pt\hbox{$>$}}}                

\newcommand{\alphas}{\alpha_{\mathrm{s}}}
\newcommand{\alphaem}{\alpha_{\mathrm{em}}}

\newcommand{\pt}{p_{\perp}}

\newcommand{\Dy}{{\Delta y}}

\renewcommand{\c}{\mathrm{c}}

\newcommand{\g}{\mathrm{g}}

\newcommand{\q}{\mathrm{q}}

\newcommand{\cbar}{\overline{\mathrm{c}}}

\newcommand{\qbar}{\overline{\mathrm{q}}}

\newcommand{\Py}{{\sc{Pythia}}~}

\begin{document}

\sloppy

\pagestyle{empty}

\begin{flushright}
LU TP 99--08 \\
May 1999
\end{flushright}
 
\vspace{\fill}

\begin{center}  \begin{Large} \begin{bf}
Drag effects in charm photoproduction\\
  \end{bf}  \end{Large}
  \vspace*{5mm}
  \begin{large}
E. Norrbin and T. Sj\"ostrand\\
  \end{large}

Department of Theoretical Physics 2, Lund University\\
S\"olvegatan 14 A, s-223 62 Lund, Sweden\\
emanuel@thep.lu.se, torbjorn@thep.lu.se\\
\end{center}

\vspace{\fill}

\begin{quotation}
\noindent
{\bf Abstract:}
We have refined a model for charm fragmentation at hadron colliders. This model
can also be applied to the photoproduction of charm. We investigate the effect of
fragmentation on the distribution of produced charm quarks. The drag effect is seen to
produce charm hadrons that are shifted in rapidity in the direction of the beam remnant. We
also study the importance of different production mechanisms such as charm in the photon
and from parton showers.
\end{quotation}
\vspace{\fill}

\clearpage
\pagestyle{plain}
\setcounter{page}{1}

In previous work \cite{previous} we studied and refined a model for the
hadronization of a low-mass string in the framework of the Lund string
fragmentation model \cite{AGIS}. The model was used to describe the leading particle
effect that has been observed at fixed-target experiments \cite{experiment}.
With a leading charmed meson defined as having the light quark in common with the incoming beam,
an asymmetry has been observed between leading and non-leading charmed
mesons, favouring leading particles in the beam fragmentation region.
In a string fragmentation framework this is understood in the following
way. Because of the colour flow in an event, the produced charm quarks normally are
colour-connected to the beam remnants of the incoming particles. This results in
the possibility for a charmed hadron to gain energy and momentum from
the beam remnant in the fragmentation process and thus be produced at a larger rapidity
than the initial charm quark. The extreme case in this direction is when the
colour singlet containing the charm quark and the beam remnant has a small invariant
mass, e.g. below or close to the two-particle threshold. Then the colour singlet,
called a {\it cluster}, will be forced to collapse into a meson, giving
a hard leading particle. The corresponding production mechanism for non-leading
particles involves sea-quarks and is therefore suppressed.

The qualitative nature of the asymmetry can thus be understood within the string
model. The quantitative predictions, however, depend on model parameters.
The model has been tuned to reproduce data on both asymmetries and single-charm
spectra at fixed target energies \cite{previous}.
Here we wish to apply the model to $\gamma$p physics
at HERA. The asymmetries are small in this case because of the higher energy
and the flavour neutral photon beam. Therefore the emphasis
will be shifted towards beam-drag effects, consequences of the photon structure
and higher-order effects.

The photon is a more complicated object than a hadron because it has two components,
one {\em direct} where the photon interacts as a whole and one {\em resolved} where it has
fluctuated into a $\q\qbar$ pair before the interaction.
This will result in very different event structures
in the two cases. This study is constrained to real photons (photoproduction) as modeled by
Schuler and Sj\"ostrand \cite{SaS} and implemented in the \Py \cite{pythia} event generator.
We include the photon flux and use cuts close to the experimental ones. We first
examine the leading-order charm spectra for direct and resolved photons,
estimate the cross section
in the two cases, and study how the fragmentation process alters the charm
spectra in the string model. Then we add some higher-order processes (flavour excitation
and quark splitting) and find that they give a significant contribution to the charm
cross section, especially for resolved photons.

\begin{figure}[htb]
\vspace*{2mm}
\mbox{
\epsfig{file=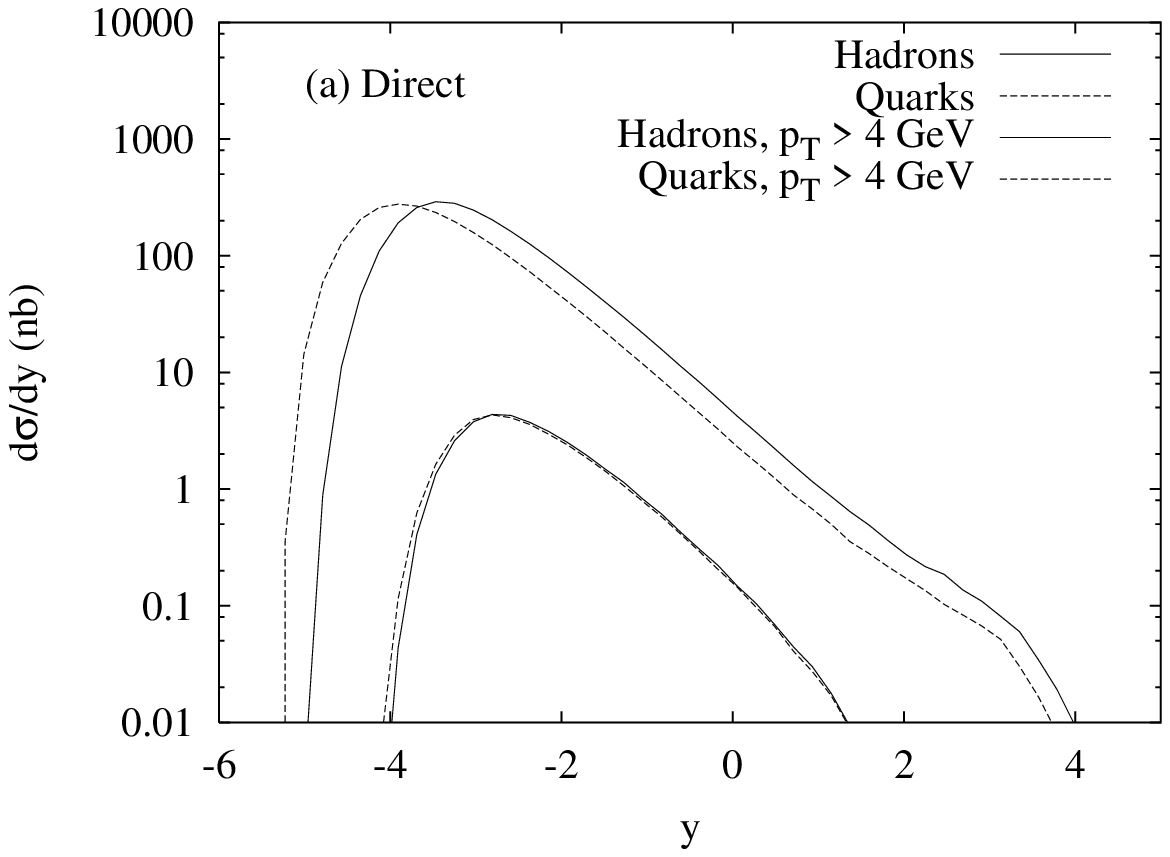, width=68mm}
\epsfig{file=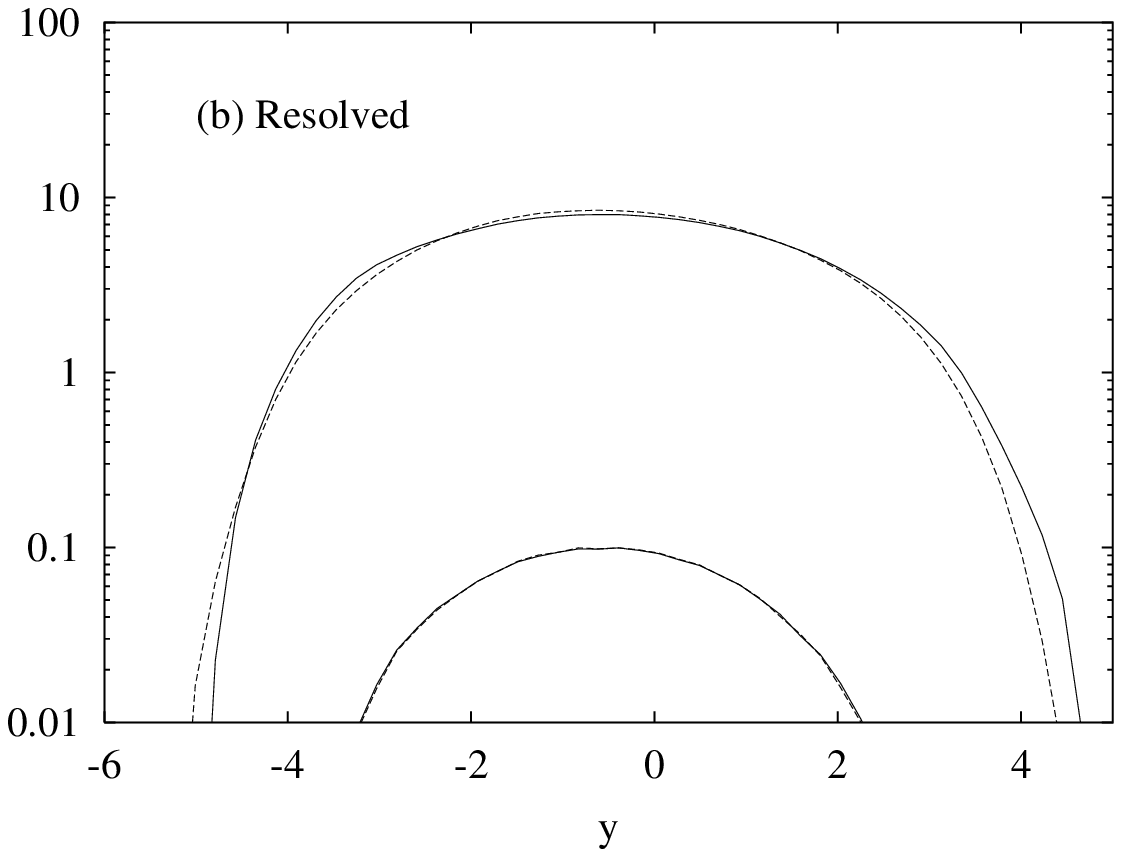, width=68mm}
}
\mbox{
\epsfig{file=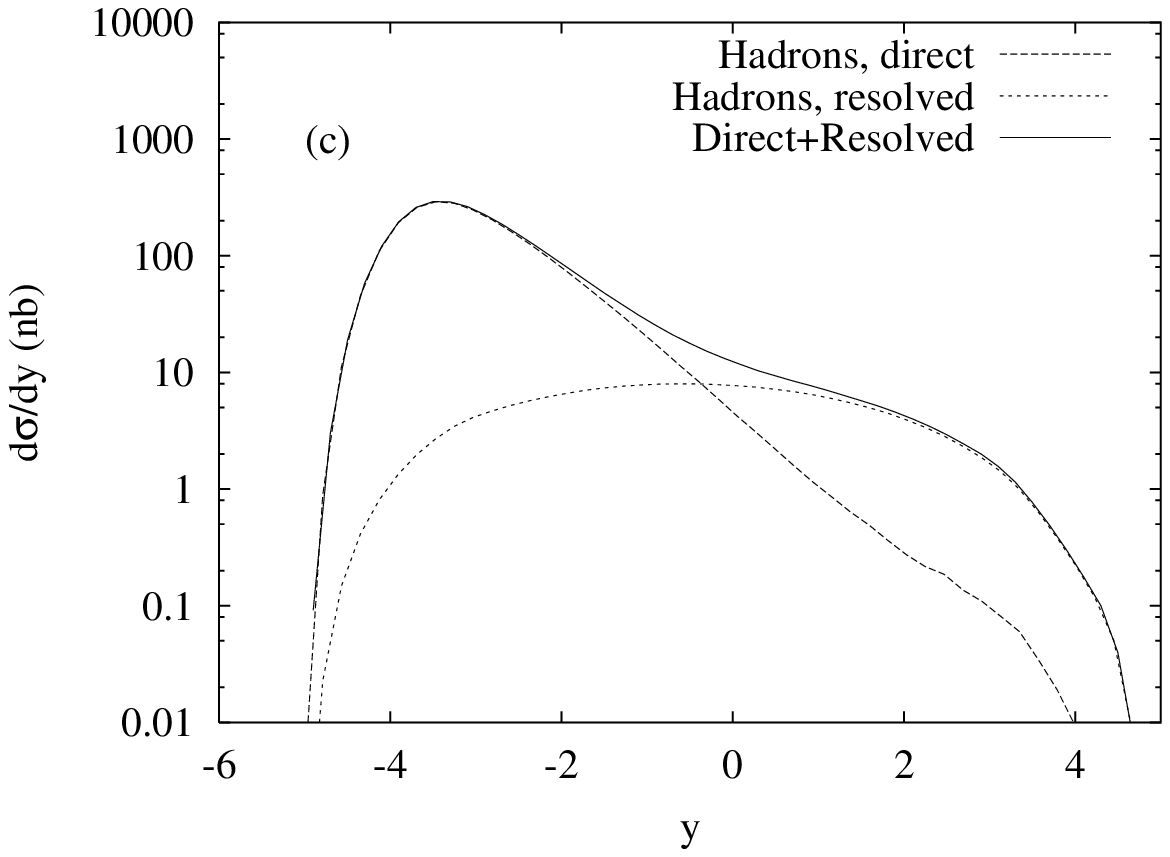, width=68mm}
\epsfig{file=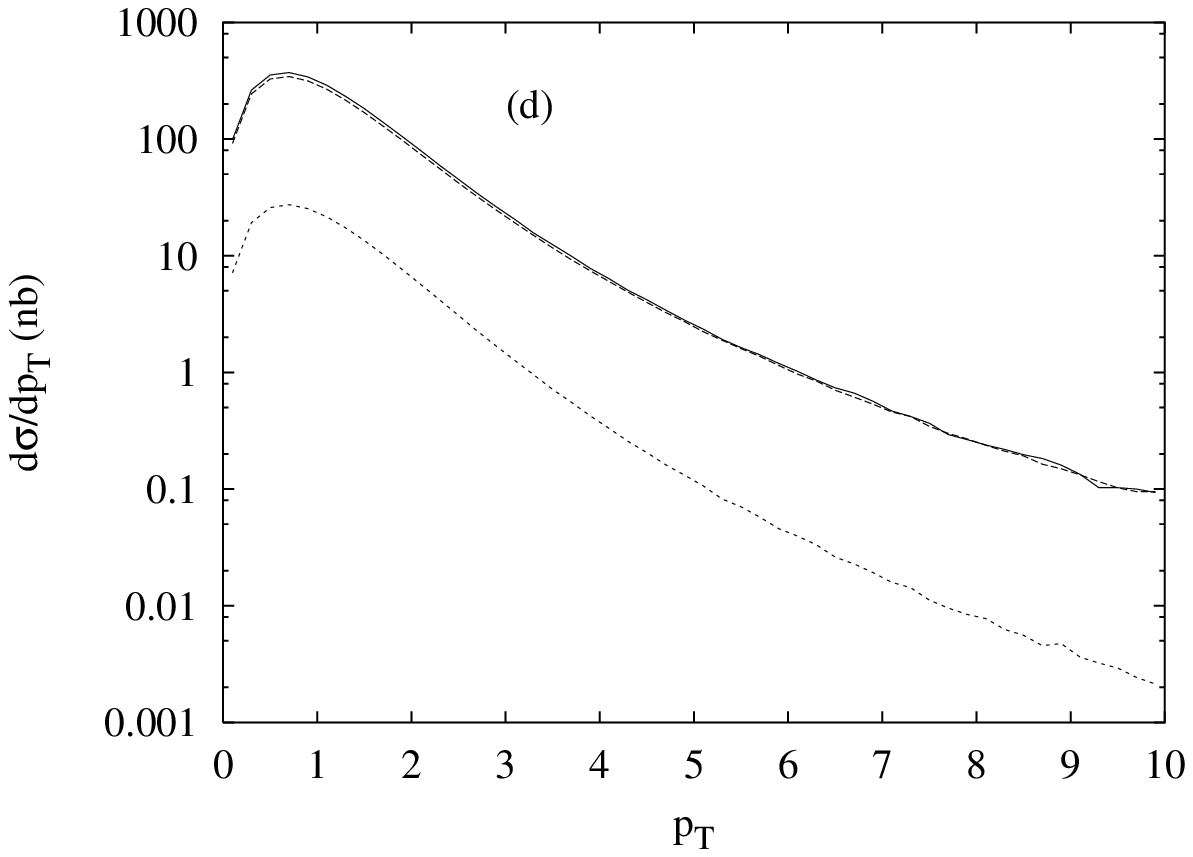, width=68mm}
}
\caption[junk]{{\it
Distribution of charmed hadrons and quarks in rapidity:
(a) direct and (b) resolved photons.
Comparison of resolved and direct processes in (c) rapidity and (d) transverse momentum.
}}
\label{LO}
\end{figure}

We consider charm photoproduction in an $\mathrm{e}^{\pm}$p collision (820 GeV protons and 27.5 GeV electrons)
with real photons ($Q^2<1$ GeV),
$130<W_{\gamma \mathrm{p}}<280$ and some different $\pt$-cuts. The analysis is done in
the $\gamma$p center of mass system using true rapidity
($y = \frac{1}{2}\ln(\frac{E+p_z}{E-p_z})$) as the main kinematical variable. The photon
(electron) beam is incident along the negative z-axis.

To leading order, the massive matrix elements producing charm are the fusion
processes $\q \gamma \rightarrow \c \cbar$ (direct), $\g \g \rightarrow \c \cbar$ and
$\q \qbar \rightarrow \c \cbar$ (resolved). Fig.~\ref{LO} shows the distribution of
charmed quarks and charmed hadrons separated into these two classes. For direct photons
the hadrons are shifted in the direction of the proton beam, since both charm quarks
are colour-connected to the proton beam remnant. In a resolved event the photon also has
a beam remnant, so the charmed hadron is shifted towards the beam remnant it is connected to.
Also note that the drag effect is a small-$\pt$ phenomenon. A jet at high $\pt$ will
not be much influenced by the beam remnant.

\begin{figure}[htb]
\vspace*{2mm}
\mbox{
\epsfig{file=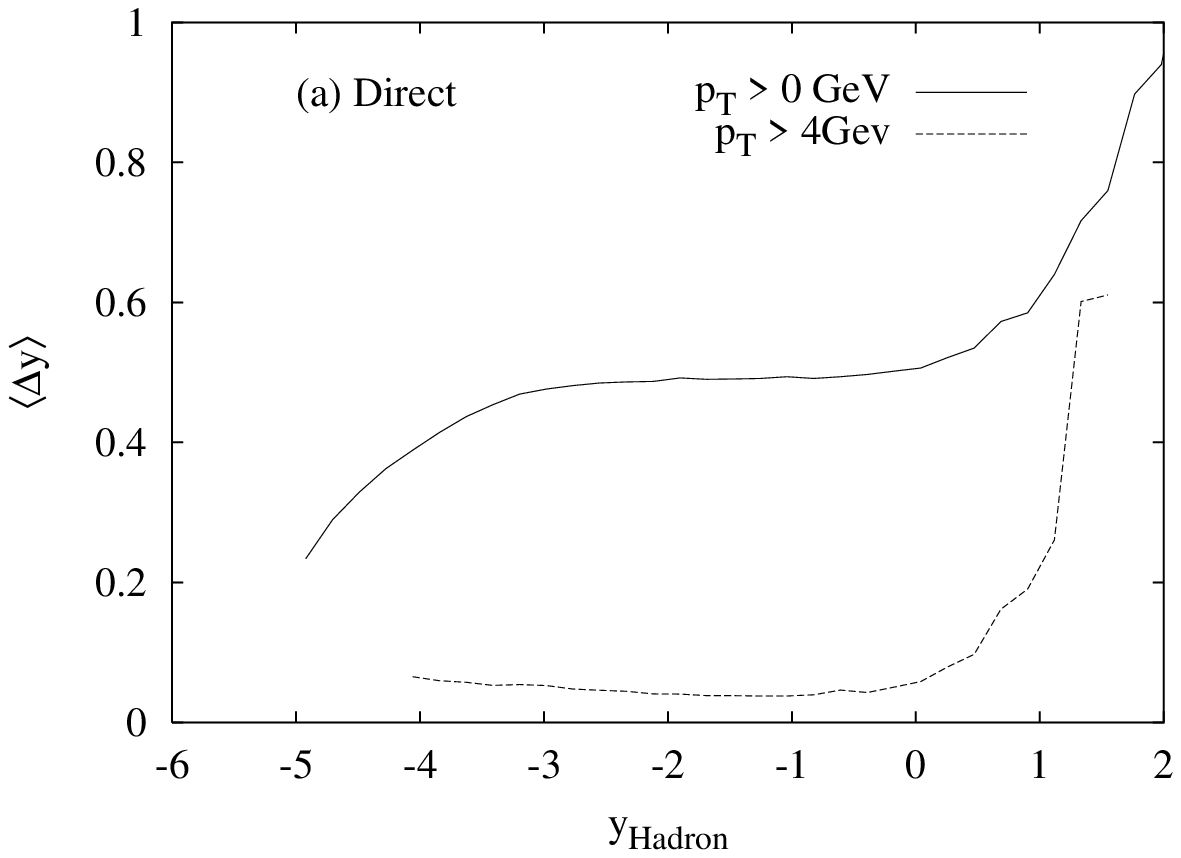, width=68mm}
\epsfig{file=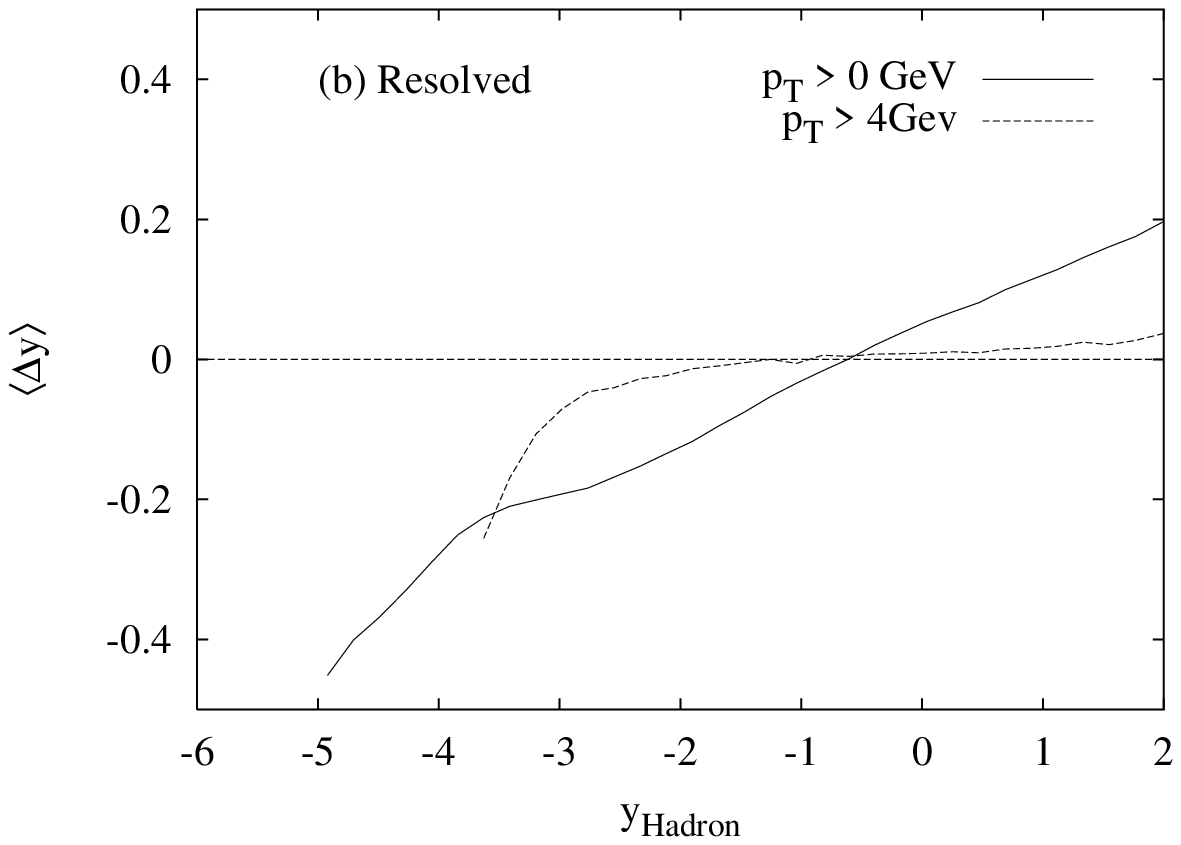, width=68mm}
}
\caption[junk]{{\it
Rapidity shift
$\langle \Dy \rangle = \langle y_\mathrm{Hadron} - y_\mathrm{Quark} \rangle$
for (a) direct and (b) resolved photons as a function of rapidity.
}}
\label{drag1}
\end{figure}

The drag effect is illustrated in Fig.~\ref{drag1} where the average rapidity shift in
the hadronization,
$\langle \Dy \rangle = \langle y_\mathrm{Hadron} - y_\mathrm{Quark} \rangle$,
is shown as a function of $y_\mathrm{Hadron}$.
For direct photons and central rapidities the shift is approximately constant. The
increasing shift for large rapidities is due to an increased
interaction between the proton remnant and the charmed quark when
their combined invariant mass is small. At large negative rapidities there is no
corresponding effect because there is no beam remnant there. The drop of
$\langle\Dy\rangle$ in this region is a pure edge effect; only those events with below-average
$\Dy$ can give a very negative $y_\mathrm{Hadron}$. For resolved photons
the shift is in the direction of the proton and photon beam remnants. Note that
what is plotted is only the mean. The width of $\Dy$ is generally larger than the
mean, so the shift can go both ways. For example the quarks at very small rapidities ($y\lsim -5$)
in Fig.~\ref{LO}b will all be shifted with $\Dy>0$ but hadrons produced there will,
on the average, come from quarks produced at larger rapidities (i.e. $\Dy<0$).
Hence the apparent contradiction with Fig.~\ref{drag1}b by these edge effects.

\begin{figure}[htb]
\vspace*{2mm}
\mbox{
\epsfig{file=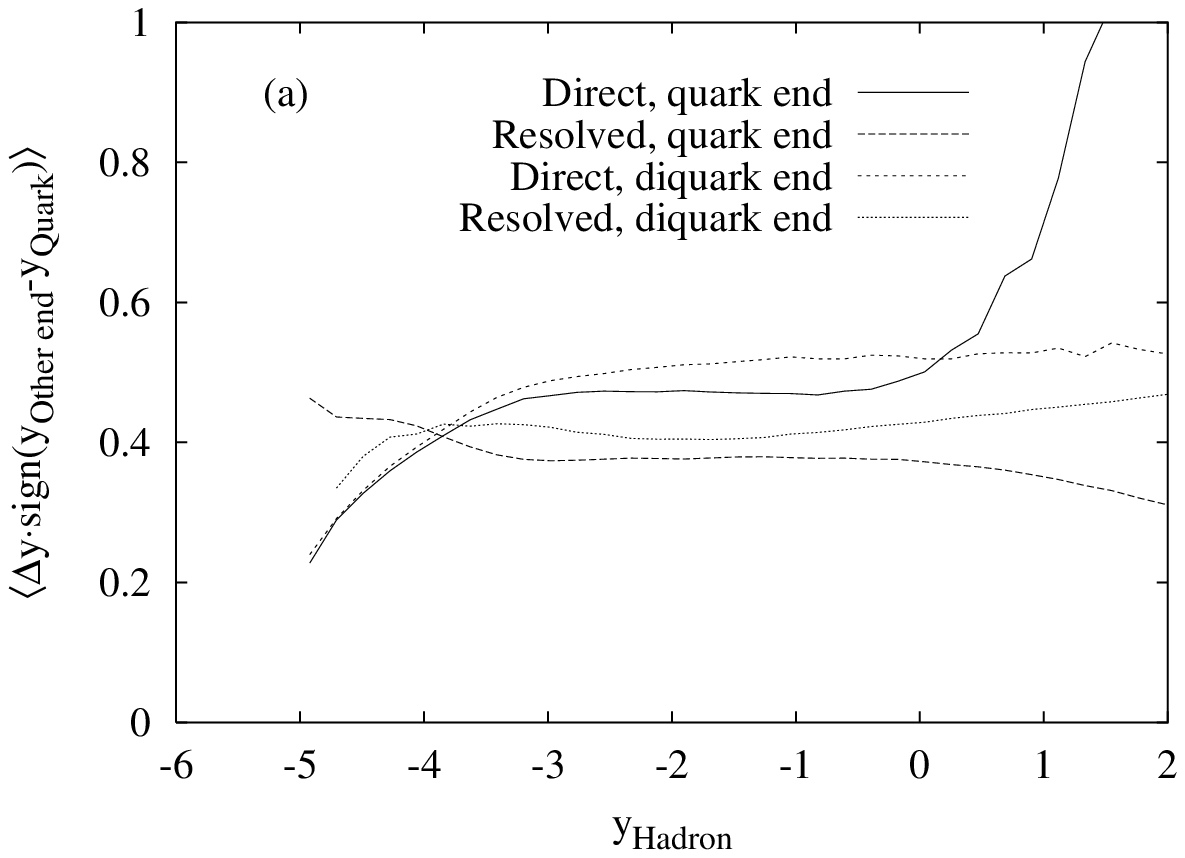, width=68mm}
\epsfig{file=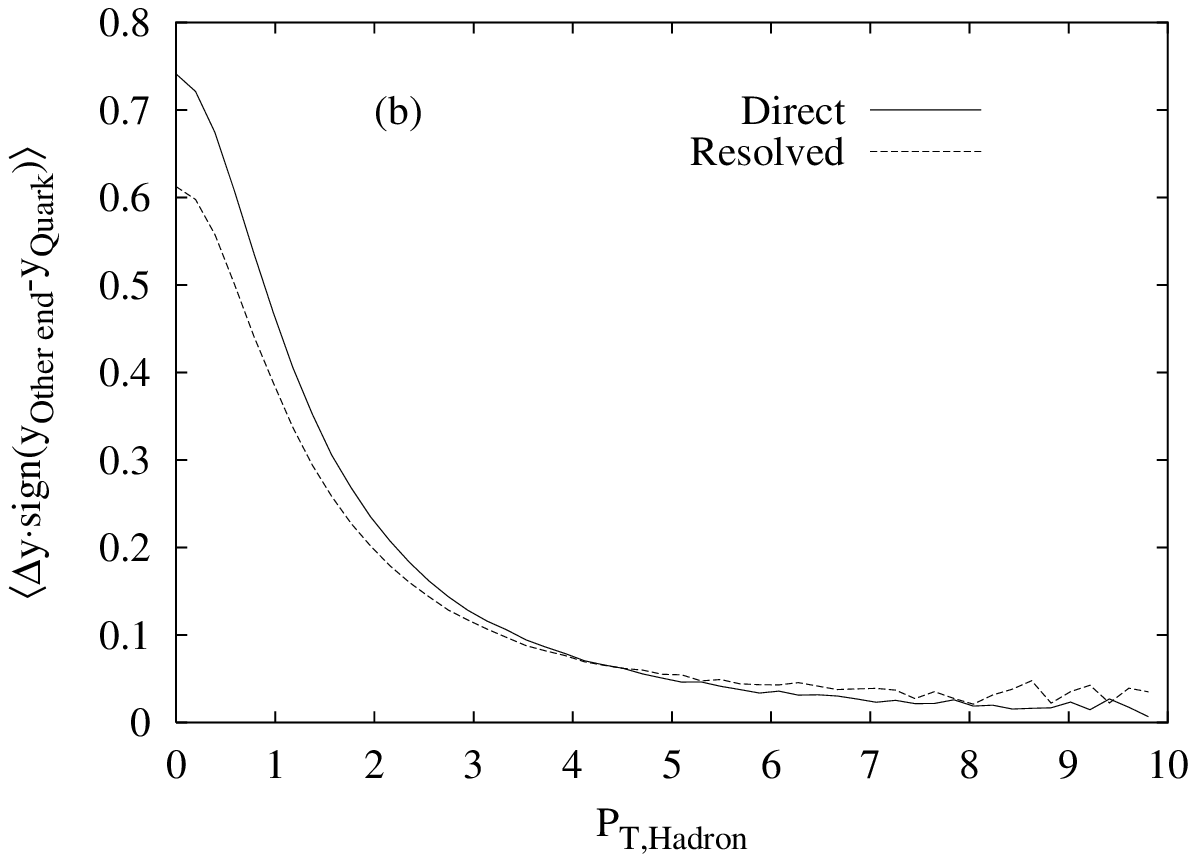, width=68mm}
}
\caption[junk]{{\it  
Rapidity shift $\langle \Dy \cdot sign(y_{Other \hspace{1mm} end} - y_{Quark}) \rangle$
as a function of (a) rapidity and (b) transverse momentum. The shift for a hadron
produced from a string containing a diquark in the other end is separated from those containing a quark.
No $\pt$ cut is applied.
}}
\label{drag2}
\end{figure}

In order to isolate the drag effect we plot the rapidity shift in the direction
of 'the other end of a string'. This is accomplished by studying
$\langle \Dy \cdot \mathrm{sign}(y_\mathrm{Other \hspace{1mm} end} - y_\mathrm{Quark}) \rangle$
as a function of $y$ and $\pt$ as shown in Fig.~\ref{drag2}.
In this case the difference between direct and resolved events are less marked,
showing the universality of string fragmentation. The remaining differences stem
from the different distributions of string masses in the two cases.

Higher-order effects can be included in an event generator through flavour excitation
(e.g. $\c \q \rightarrow \c \q$) and parton showers (gluon splitting, $\g \rightarrow
\c \cbar$). This approach is in some ways complementary to full NLO calculations.
A NLO calculation of the charm cross section contains all diagrams up to order $\alphas^3$
($\alphas^2\alphaem$ for direct photons)
whereas a Monte Carlo event generator simulating parton showers/flavour excitation
contains all diagrams of order
$\alphas^2$ ($\alphas \alphaem$ for direct photons) and an approximation to all higher orders.
In this way some processes that
are not included in a NLO calculation are approximated. Some examples are
$\gamma \q\rightarrow \c\cbar\q\g$, $\g\gamma \rightarrow \q\qbar\c\cbar$
and $\g\g \rightarrow \c\cbar\c\cbar$.

\begin{figure}[htb]
\vspace*{2mm}
\mbox{
\epsfig{file=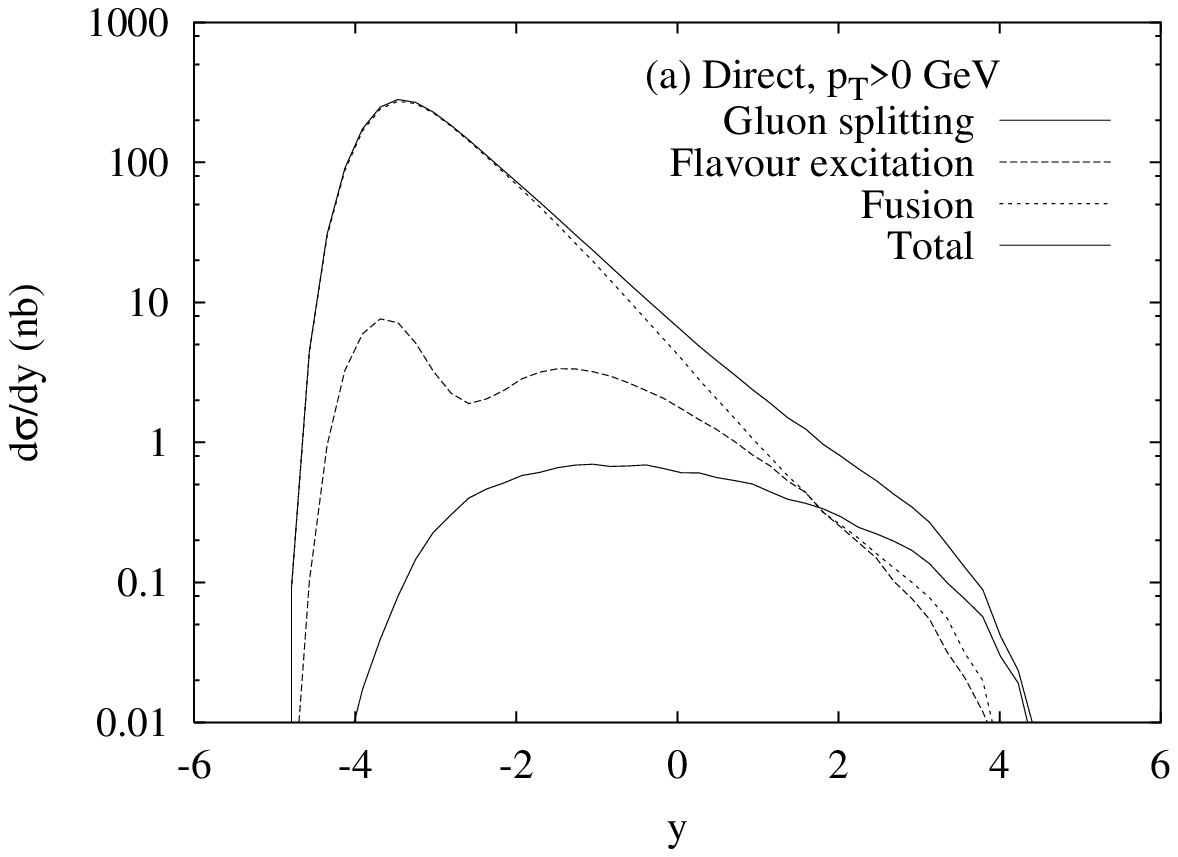, width=68mm}
\epsfig{file=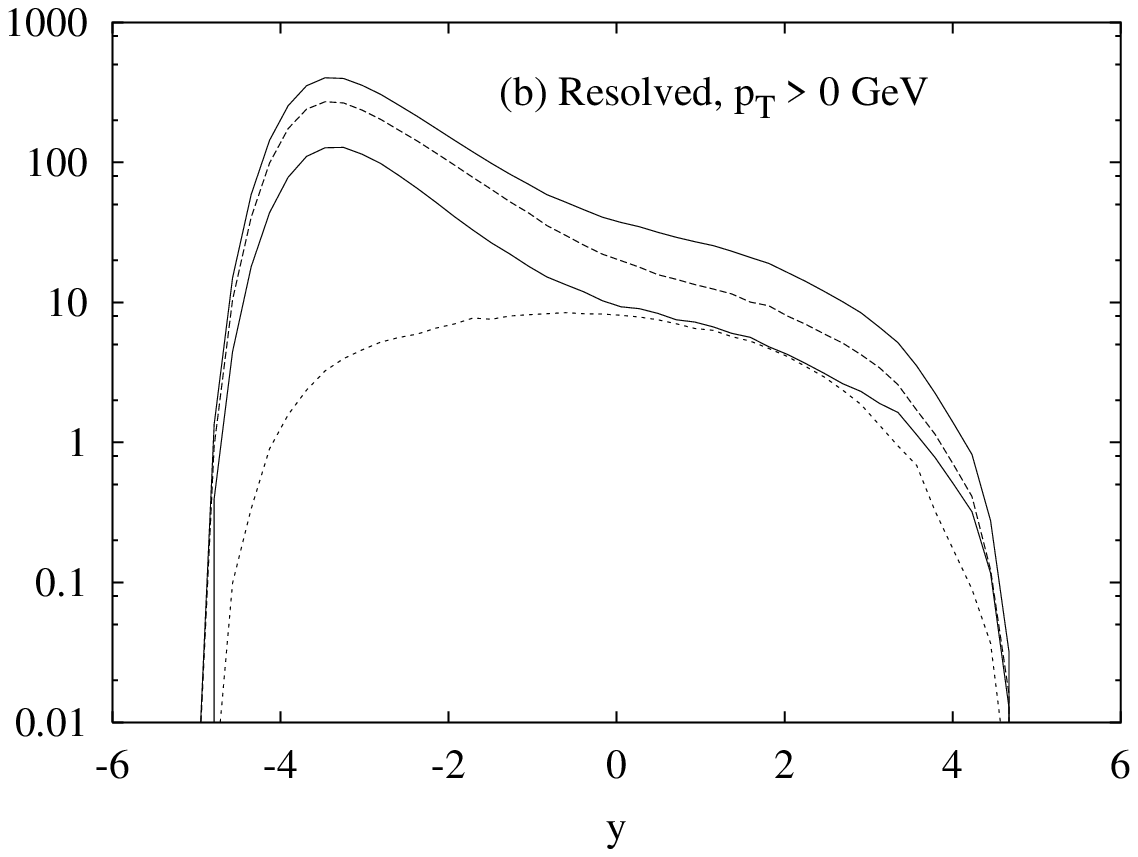, width=68mm}
}
\mbox{
\epsfig{file=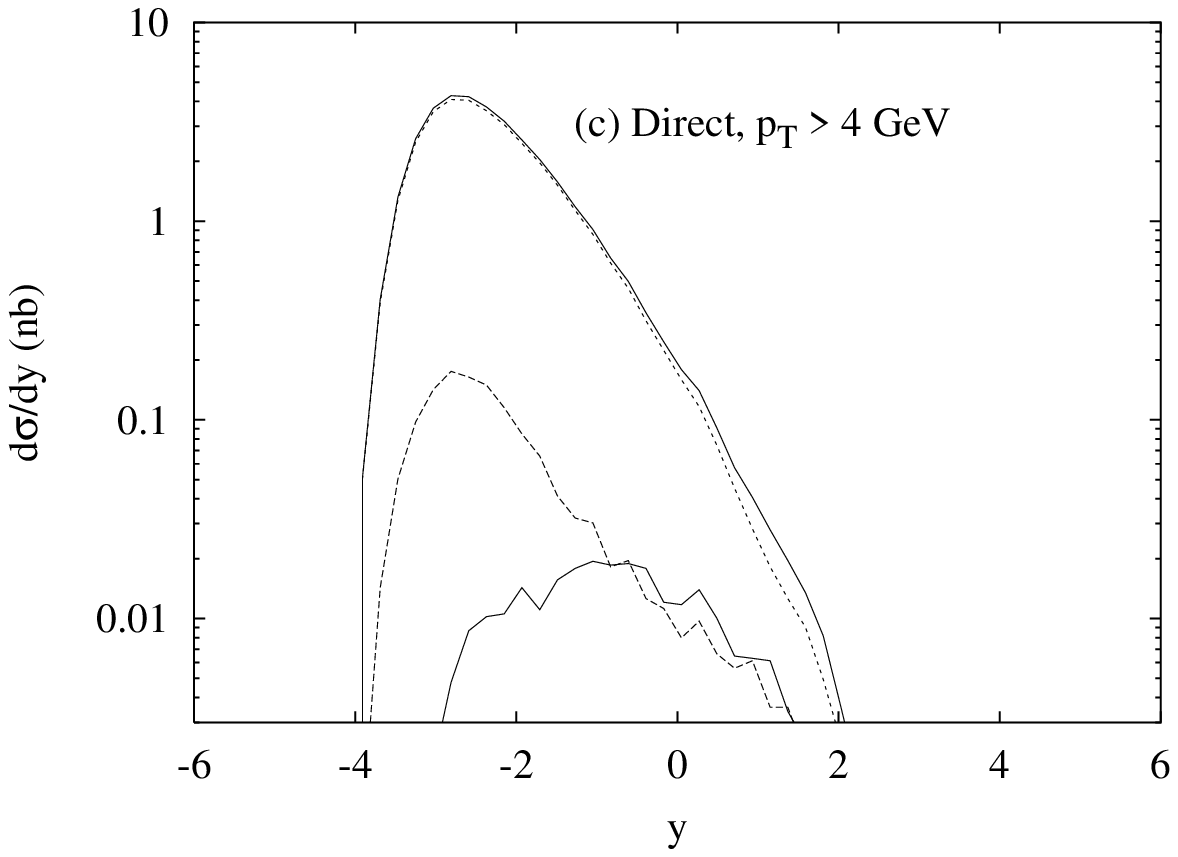, width=68mm}
\epsfig{file=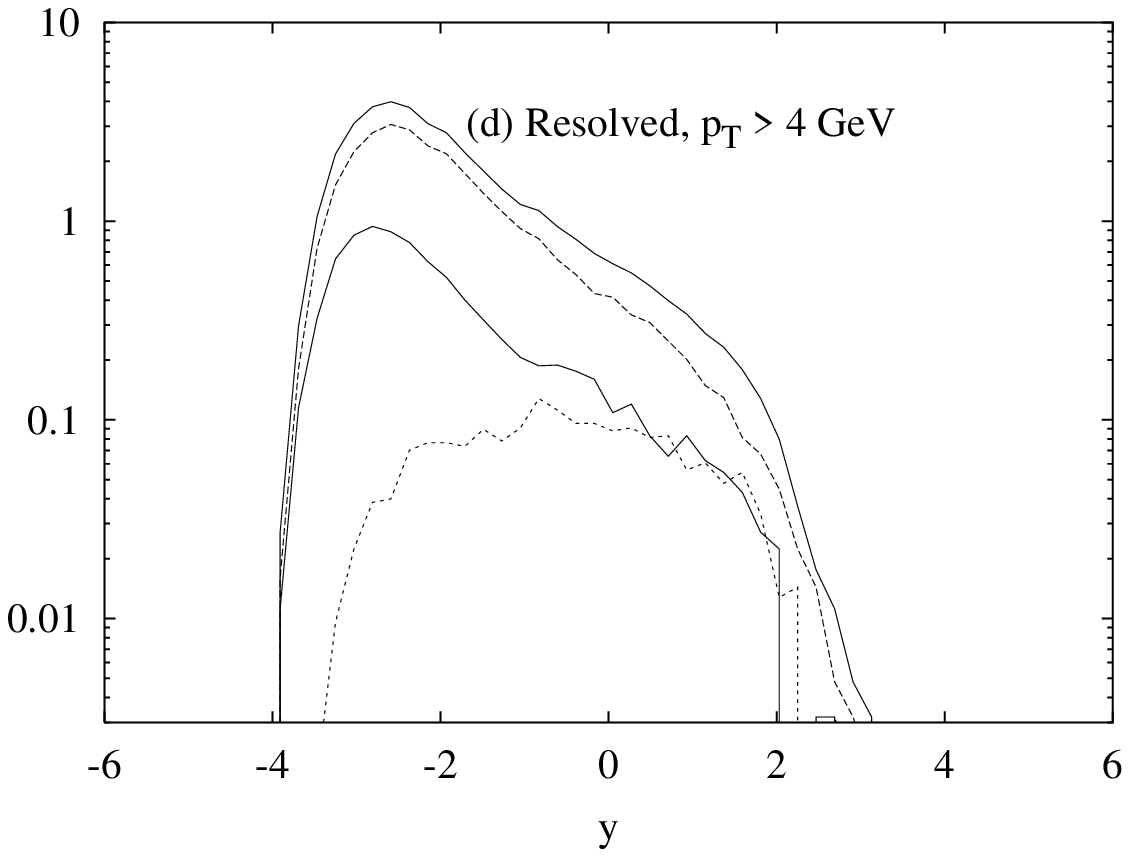, width=68mm}
}
\mbox{
\epsfig{file=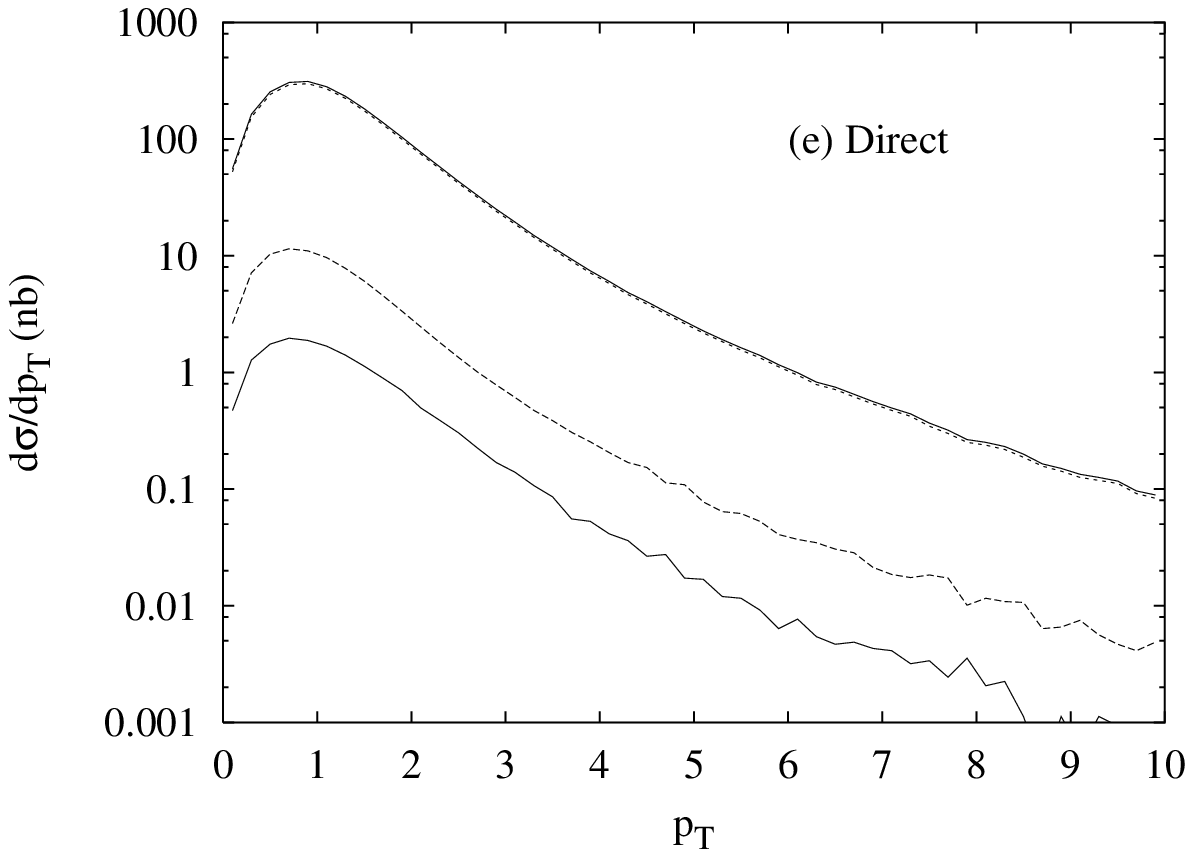, width=68mm}
\epsfig{file=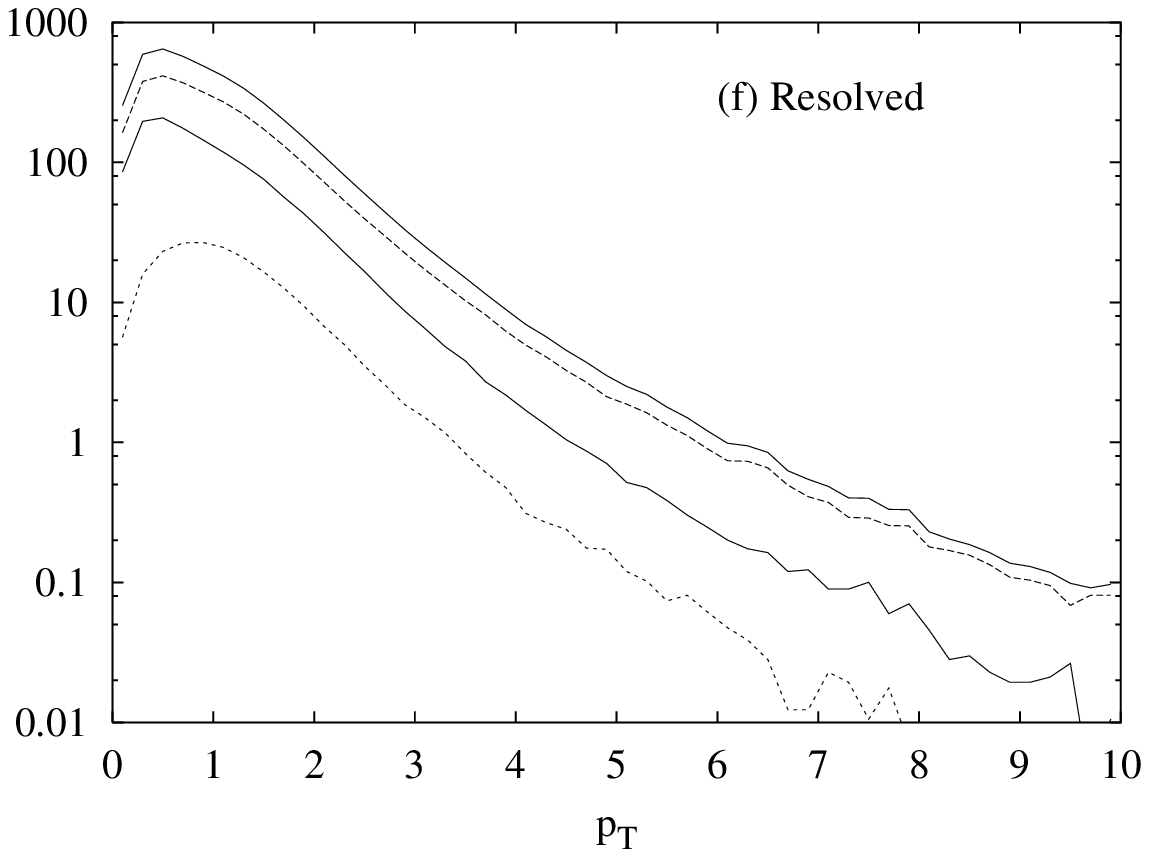, width=68mm}
}
\mbox{
\epsfig{file=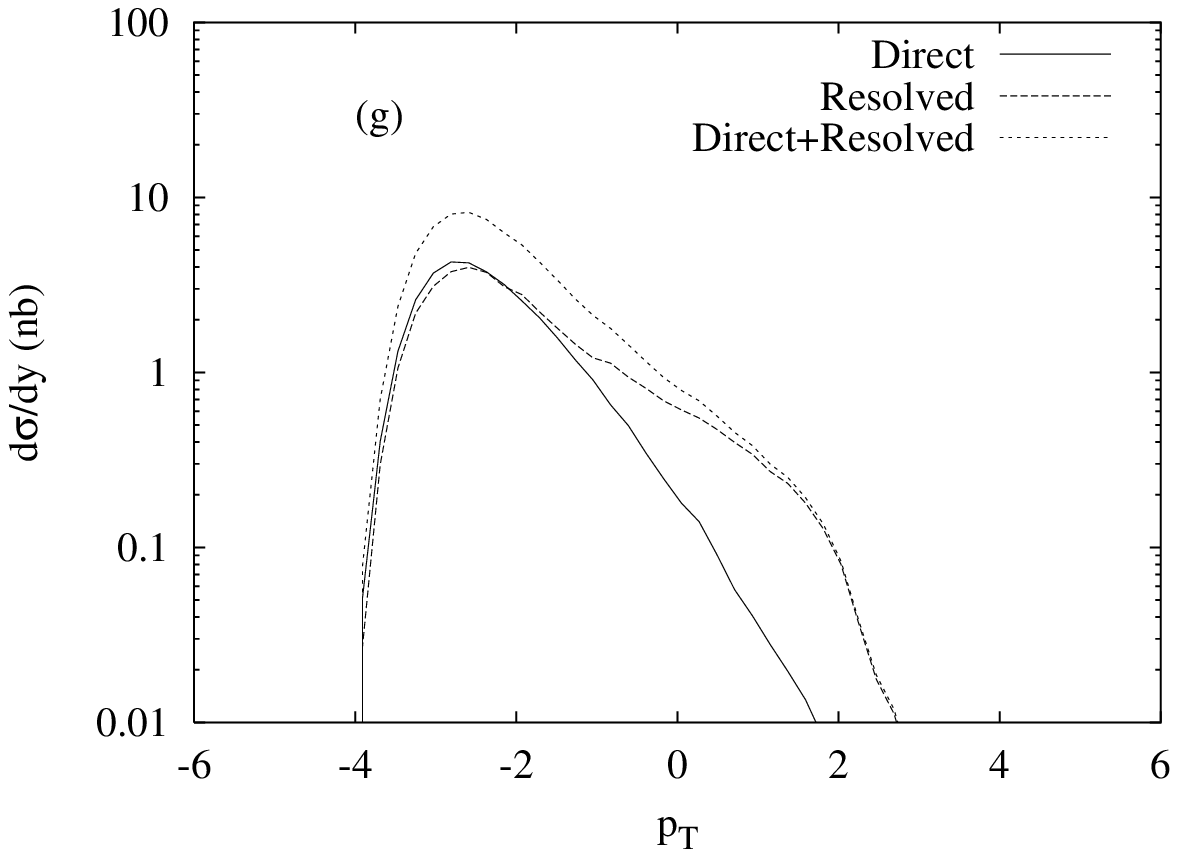, width=68mm}
\epsfig{file=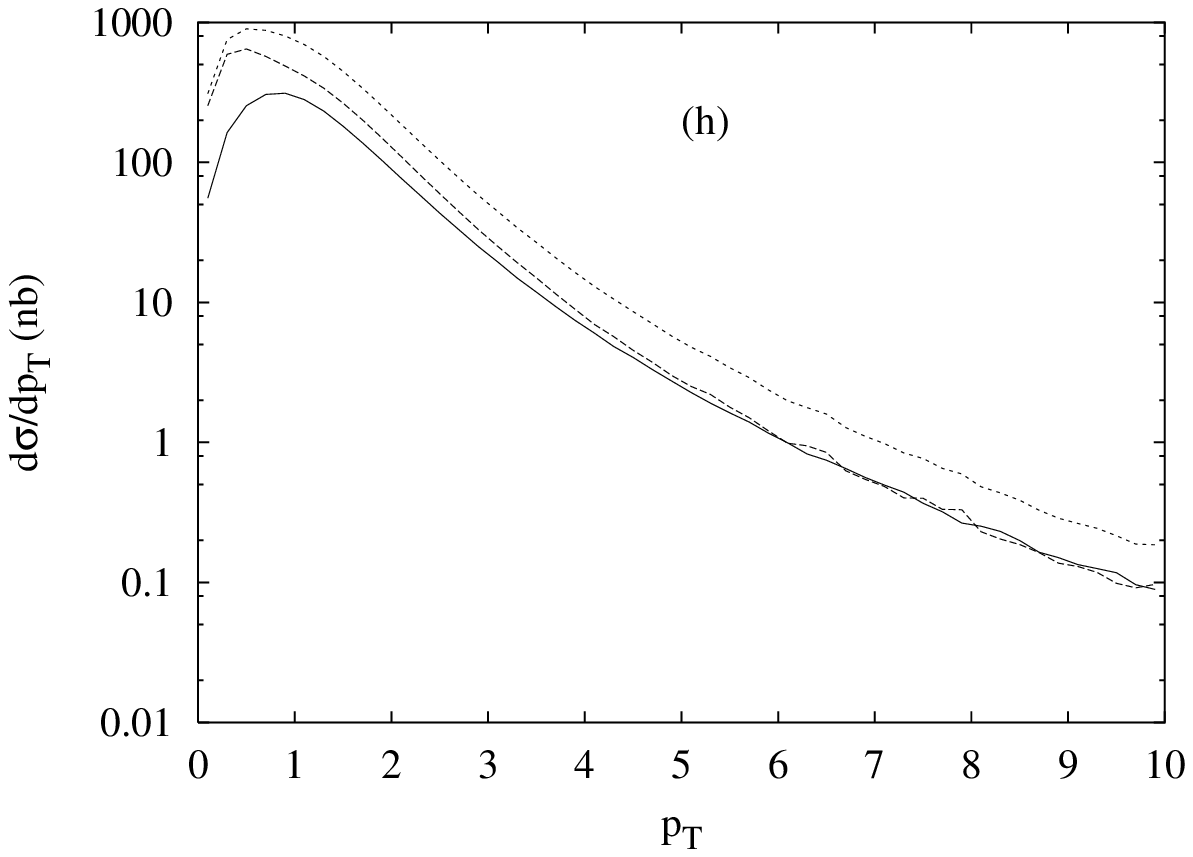, width=68mm}
}
\caption[junk]{{\it
The cross section divided into different production mechanisms and different photon structure.
(a) Direct and (b) resolved photons with $\pt > 0$ \textup{GeV}.
(c) Direct and (d) resolved photons with $\pt > 4$ \textup{GeV}.
(e) Direct and (f) resolved photons in $\pt$.
We add the components together for (g) rapidity ($\pt > 4$ \textup{GeV}) and (h) transverse momentum.
}}
\label{HO}
\end{figure}

At HERA energies, higher-order effects give large contributions to the cross section.
In Fig.~\ref{HO} the cross section is divided into different production channels
for direct and resolved photons. We note that now the cross sections are of the same
order of magnitude and the major contribution
in the resolved case is flavour excitation. The details of course depend
on the parameterization of the photon structure.

The double peak structure in the flavour excitation process for direct photons
is because the charm quark in the beam remnant at low $\pt$ is also included.
This peak disappears when a $\pt$ cut is introduced (Fig.~\ref{HO}c). 

\begin{figure}[htb]
\vspace*{2mm}
\mbox{
\epsfig{file=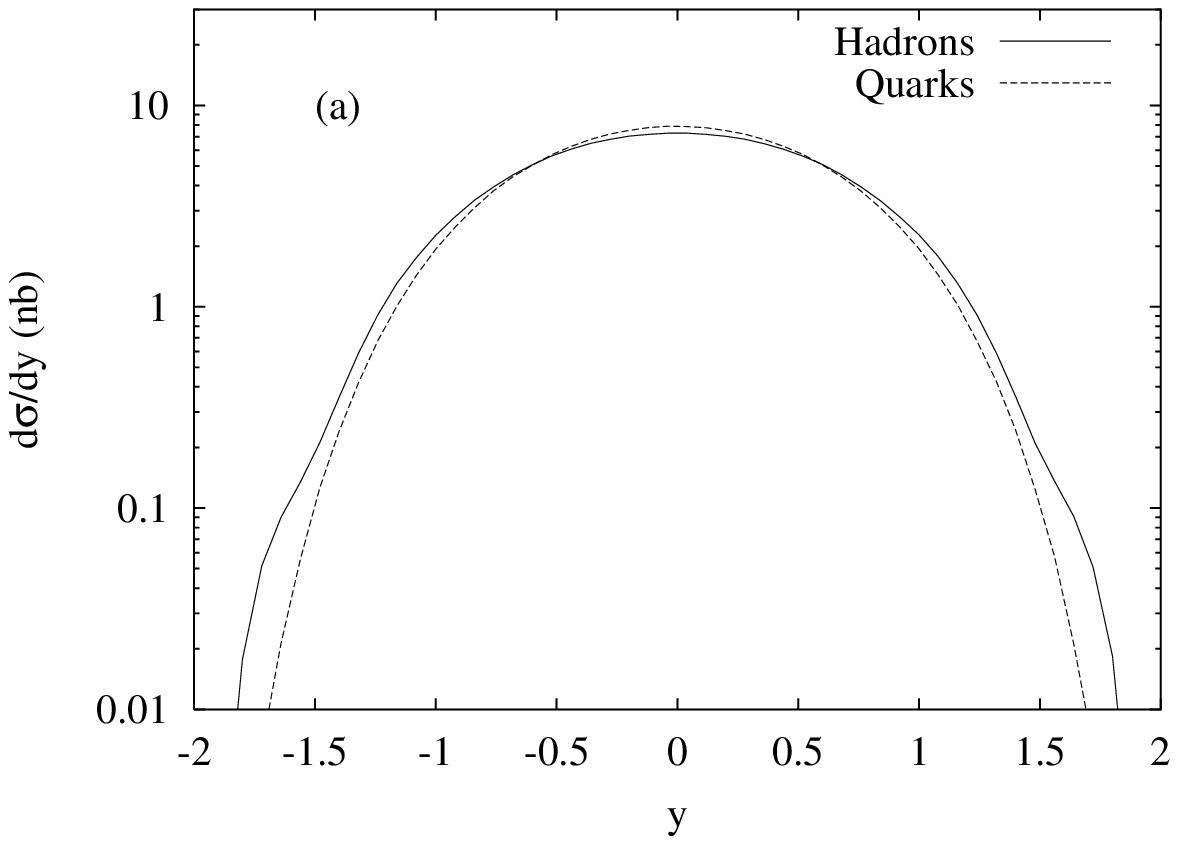, width=68mm}
\epsfig{file=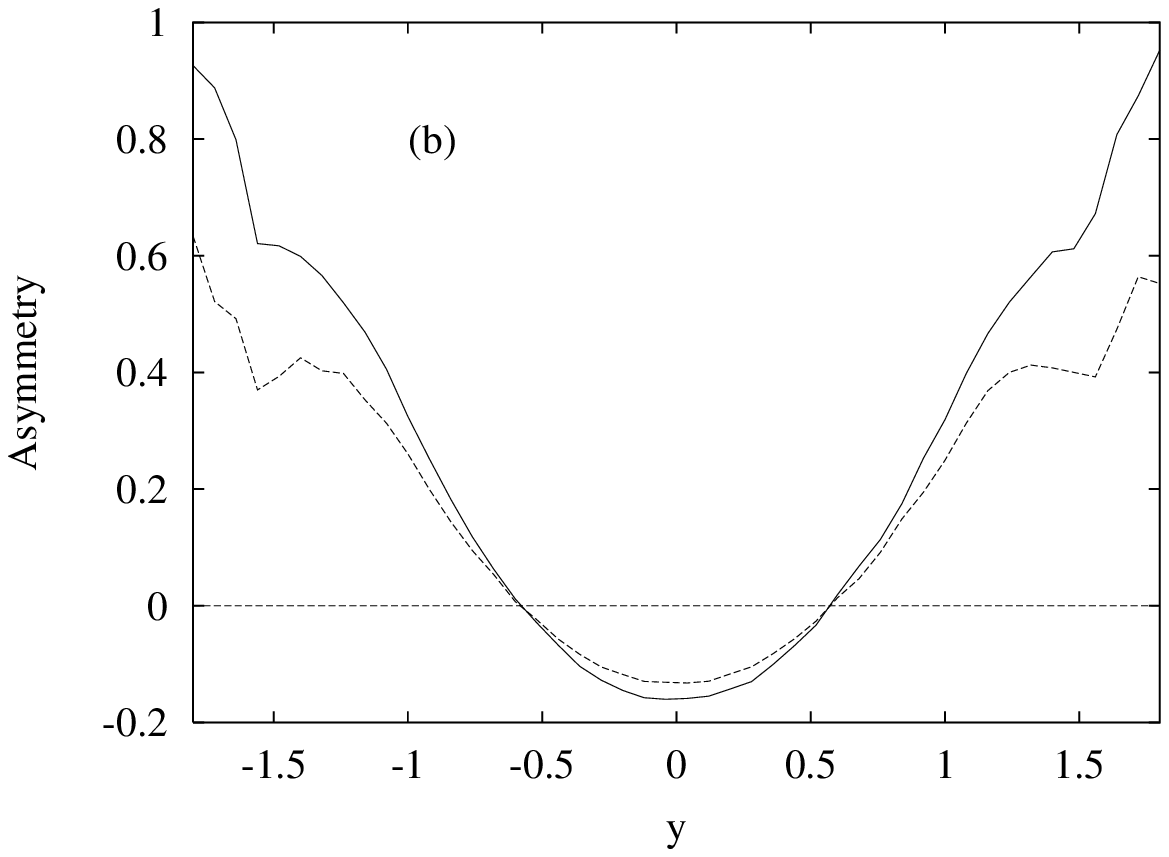, width=68mm}
}
\caption[junk]{{\it
(a) Distribution of bottom quarks and hadrons at HERA-B (pp at 40 \textup{GeV} CM-energy, i.e. no nuclear target effects included).
(b) Asymmetry ($A=(\sigma(\overline{\mathrm{B}^0})-\sigma(\mathrm{B}^0))/(\sigma(\overline{\mathrm{B}^0})+\sigma(\mathrm{B}^0))$
as a function of rapidity for two parameterizations of the beam remnant distributions
(full=uneven, dashed=even) \cite{previous}. This asymmetry is due to drag effects where hadrons
containing bottom quarks have been shifted more towards the beam remnant than those containing anti-quarks.
The diquark with, on the average, larger energy/momentum than the quark of the proton beam remnant, is colour
connected to the bottom quark, thus shifting it more than the anti-quark.
}}
\label{HERAB}
\end{figure}

The physics discussed here has
consequences also for b-production at HERA-B. Because of the larger mass of the b-quark,
drag/collapse effects are expected to be smaller. However, this is compensated by
the smaller CM-energy when the HERA proton beam is used on a fixed target,
giving non-negligible effects as shown in Fig.~\ref{HERAB}.
An understanding of these aspects are important when studying CP violation
in the $\mathrm{B}^0\overline{\mathrm{B^0}}$ system \cite{CP}.

In summary we have improved the modelling of charm in the \Py event generator by a
consideration of charm hadroproduction data \cite{previous}. In this note we study
beam-drag effects at HERA and it should be interesting to look for experimental
signatures, e.g. differences between NLO and data. We also show that higher order
effects give important contributions to the charm production spectra at HERA energies,
especially for resolved photons.

\end{document}